\documentclass[a4paper,12pt]{article}
\usepackage{fancyhdr}

\usepackage{amssymb}
\usepackage{bm}

\newcommand{\comma}{\ , \quad}

\begin{document}

\title{
Couplings and spectra in modular inflation
}

\author{
Wan-Il Park\footnote{wipark@muon.kaist.ac.kr} \
and
Ewan D. Stewart
\\[1ex] \textit{
Department of Physics, KAIST, Daejeon 305-701, South Korea
}}

\maketitle

\abstract{
We analyze how the spectrum of perturbations produced in a multi-component modular inflation model proposed by Kadota and Stewart depends on couplings between the two moduli.
Although some simple direct couplings give essentially the same results as the original model, $dn/d\ln k \propto n-1$, simple indirect couplings produce a power law spectrum, $n-1 = \textrm{constant}$, which can naturally be close to scale invariant.
}

\thispagestyle{fancy}
\rhead{KAIST-TH 2007/01}

\section{Introduction}

In string/M theory there are field directions, called moduli, which are exactly flat before taking into account supersymmetry breaking.
The potential generated by supersymmetry breaking still allows Planckian field values at low energy.
This property has attracted much attention to their cosmology \cite{ModuliProblem,ModularInflation,flatening}.

The cosmology of moduli has been challenging.
For example, stabilizing the moduli can be difficult \cite{MDNS}, and even if there are stable points on the moduli potential, the cosmological evolution may not lead to those points \cite{RBPJS}.
Additionally, large amplitude coherent moduli oscillations are typically generated after inflation, with potentially disasterous consequences \cite{ModuliProblem}.

On the other hand, inflation can be realized simply and naturally by a modulus rolling off a maximum of its potential.
This has been called modular inflation \cite{ModularInflation,flatening}.
Typically, one has eternal inflation at the maximum, and this can provide a relatively safe initial condition for the observable evolution of the universe, because it occurs at the scale where the moduli are stabilized.

Furthermore, any excess moduli can be safely diluted by thermal inflation \cite{DHLEDS}, and although the usual scenarios of baryogenesis tend not to be consistent with thermal inflation, baryogenesis can occur very naturally at the end of thermal inflation \cite{DJKKWIPEDS}.

However, the usual simple models of modular inflation are not satisfactory, since the inflaton mass would be expected to be of the order of the Hubble parameter, so the inflation does not produce scale invariant perturbations without fine tuning.
To see if this difficulty is based on over simplified models, one should consider the nature of moduli more carefully.

The moduli field space has points of enhanced gauge symmetry, separated by Planckian field values.
The symmetries tend to force these points to be extrema of the moduli potential after supersymmetry is broken, and thus these points are likely to be important for inflation.
The symmetries also tend to enforce approximate global symmetries when the moduli potential is Taylor expanded about these points.
These approximately flat angular directions may also be important for inflation.
Furthermore, the symmetries also tend to force these points to be the points where other fields couple to the moduli, i.e.\ when the moduli are at these points, other fields with unsuppressed couplings to the moduli can be light.
These couplings generate field dependent renormalization of the moduli potential, which may also be important for inflation.

At Planckian distances from the points of enhanced symmetry, the general form of the moduli potential is
\begin{equation} \label{genpot}
V(\bm\Phi) = M^4_\mathrm{s} \mathinner{F(\bm\Phi / M_\mathrm{Pl})}
\end{equation}
where $\bm\Phi$ is a point in the complex moduli space, $M_\mathrm{s}$ is the scale of supersymmetry breaking, $F$ is a function with order unity dimensionless coefficients, and $M_\mathrm{Pl} = 1 / \sqrt{8 \pi G}$.
Note that the moduli space is multi-dimensional, so we expect some couplings, with coefficient $M^4_\mathrm{s}$, between the different field directions.

Nearer a point of enhanced symmetry, $\bm\Phi_i$, the moduli potential feels the effect of the couplings at the point of enhanced symmetry, which generate extra field dependent renormalisation terms in the moduli potential of the form
\begin{equation} \label{genrenorm}
\alpha m_\mathrm{s}^2 \left\| \bm\Phi - \bm\Phi_i \right\|_1^2 \ln \left( \left\| \bm\Phi - \bm\Phi_i \right\|_2^2 + m_\mathrm{s}^2 \right) + \ldots
\end{equation}
where $\alpha$ is a standard renormalisation coefficient depending on the couplings, $m_\mathrm{s}^2 = M_\mathrm{s}^4 / M_\mathrm{Pl}^2$ is the soft supersymmetry breaking mass scale, and the precise definition of the norms, $\|\cdots\|_1^2$ and $\|\cdots\|_2^2$, also depends on the details of the couplings.

Recently, Kadota and Stewart took into account the above features of the moduli potential and proposed a successful scenario of modular inflation \cite{KKEDS,KKEDS2}.
One of their models has a two complex dimensional moduli space, parameterized by the fields $\Psi$ and $\Phi$.
The potential is assumed to have a simple maximum in the $\Psi$ direction at a point of enhanced symmetry, and that $\Psi$ rolls off from this maximum.
Along the $\Phi$ direction, the potential is assumed to be stable when $\Psi$ is small and to become unstable as $\Psi$ rolls away due to couplings between $\Psi$ and $\Phi$.
Additionally, it is assumed that for small $\Psi$, the field dependent renormalization effects discussed above are strong enough to change the sign of $\Phi$'s effective mass before being cut off by the soft supersymmetry breaking mass scale.
This turns the local minimum along the $\Phi$ direction at the point of enhanced symmetry into a local maximum, with the local minimum shifted out to some field value typically exponentially smaller than $M_\mathrm{Pl}$ but exponentially larger than $m_\mathrm{s}$.
If $\Phi$ has greater than $\mathbb{Z}_2$ symmetry about the point of enhanced symmetry, then this shifted minimum is approximately flat in the angular direction.
Then the angular fluctuations are approximately scale invariant and will be magnified as $\Phi$ rolls away after being released by $\Psi$.
In this way, they were able to obtain approximately scale invariant final curvature perturbations with spectral index
\begin{equation}
n - 1 = - A k^{2\nu}
\end{equation}
with $\nu \sim 1$.
This can fit the current observations well \cite{KAKKEDS}, with a distinctive observational signature on small scales.

However, even this model is over simplified since they did not consider the likely couplings between the moduli.
Motivated by this, we investigate how the spectrum depends on various additional couplings that might be expected in such a model.

In Section~\ref{model}, we describe briefly the original model proposed in Ref.~\cite{KKEDS} and explain our additional couplings that might typically be expected in such a model.
In Section~\ref{background}, we describe the homogeneous background evolution of the fields, and in Section~\ref{perturbation}, we calculate the perturbations.
Finally, our conclusion is given in Section~\ref{conclusion}.

\section{Model}
\label{model}

Following Ref.~\cite{KKEDS}, we consider two complex moduli with potential of the form
\begin{equation}
V = V(\Phi,\Psi)
\end{equation}
with \cite{CLLSW}
\begin{equation}
\frac{\partial^2 V}{\partial\Phi\partial\Phi^*} \sim \frac{\partial^2 V}{\partial\Psi\partial\Psi^*} \sim \frac{V}{M_\mathrm{Pl}^2}
\end{equation}
We will set $M_\mathrm{Pl} = 1$ throughout most of the paper.
We assume that both $\Phi$ and $\Psi$ are initially near extrema of their potentials
\begin{equation}
\left| \frac{\partial V}{\partial\Phi} \right| \ll V
\comma
\left| \frac{\partial V}{\partial\Psi} \right| \ll V
\end{equation}
and set $\Phi = \Psi = 0$ at those extrema, so initially $|\Phi| \ll 1$ and $|\Psi| \ll 1$.

We assume that $\Phi$ has some symmetry about $\Phi = 0$ greater than $\mathbb{Z}_2$, so that the leading dependence of the potential on $\Phi$ will be $|\Phi|^2$.
Decomposing $\Phi$ into its radial and angular parts, $\Phi = \phi e^{i\theta} / \sqrt{2}$, we have in mind that fluctuations in $\theta$ will dominate the final curvature perturbation \cite{MSEDS,KKEDS}.
Taking into account the $\phi$-dependent renormalisation of $\Phi$'s mass, generated by $\Phi$'s gauge and Yukawa couplings at the point of enhanced symmetry $\Phi=0$, i.e.\ terms of the form of Eq.~(\ref{genrenorm}), the purely $\Phi$-dependent part of the potential is
\begin{equation}\label{Vphirough}
V(\Phi) \sim V_0 + \frac{1}{2} m^2_\phi \left( 1 + \alpha_\phi \ln{\phi} + \ldots \right) \phi^2 + \dots
\end{equation}
We assume $m^2_\phi$ and $\alpha_\phi$ are positive.
We expect $m^2_\phi \sim V_0 \sim M^4_\mathrm{s}$ and $\alpha_\phi$ to be small but not very small, since it is a standard renormalisation coefficient depending on $\Phi$'s gauge and Yukawa couplings.
We assume that the renormalisation is strong enough to change the sign of $\partial^2 V / \partial\phi^2$ before the renormalisation is cut off at $\phi^2 \sim m^2_\mathrm{s}$, generating a non-trivial minimum at $\phi = \phi_* \sim e^{-1/\alpha_\phi} M_\mathrm{Pl}$.
In the neighborhood of this minimum
\begin{equation} \label{Vphilead}
V(\Phi) = \frac{1}{2} m^2_\phi \left\{ - \frac{1}{2} \alpha_\phi + \alpha_\phi \ln{\left(\frac{\phi}{\phi_*}\right)} + \mathcal{O} \left[ \alpha_\phi^2 \ln^2{\left(\frac{\phi}{\phi_*}\right)} \right] \right\} \phi^2 + \ldots
\end{equation}
Note that this equation is precise while Eq.~(\ref{Vphirough}) is heuristic.
See Ref.~\cite{flatening} for a more complete treatment.

We assume that the extremum at $\Psi = 0$ is a maximum.
Writing $|\Psi| = \psi / \sqrt{2}$, the purely $\Psi$-dependent part of the potential is
\begin{equation}\label{Vpsi}
V(\Psi) = V_0 - \frac{1}{2} m^2_{\psi} \psi^2 + \ldots
\end{equation}
where $m^2_\psi \sim V_0 \sim M^4_s$ and $m^2_\psi$ may depend on the phase of $\Psi$.
The phase of $\Psi$ will probably be constant while cosmologically relevant scales leave the horizon, so we suppress any dependence on the phase of $\Psi$ here and in the following.

The full potential is
\begin{equation}\label{Vfull}
V = V_0 + \frac{1}{2} m^2_\phi \left[ - \frac{1}{2} \alpha_\phi + \alpha_\phi \ln\left(\frac{\phi}{\phi_*}\right) \right] \phi^2 - \frac{1}{2} m^2_{\psi} \psi^2 + V_\mathrm{int}(\Phi,\Psi) + \ldots
\end{equation}
where $V_\mathrm{int}(\Phi,\Psi)$ contains interactions between $\Phi$ and $\Psi$.
We consider various possibilities for these interactions.

\subsection{Direct couplings}

These couplings come from the potential in Eq.~(\ref{genpot}).

\paragraph{Linear coupling}

If $\Psi$ has no symmetry and $\psi$ is small, one would expect the leading coupling of $\Psi$ to $\Phi$ to be linear
\begin{equation} \label{case1}
V_\mathrm{int}(\Phi,\Psi) = - \frac{1}{2} \lambda_1 m^2_\phi \psi \phi^2
\end{equation}

\paragraph{Quadratic coupling}

If $\Psi$ has symmetry and $\psi$ is small, one would expect the leading coupling of $\Psi$ to $\Phi$ to be quadratic
\begin{equation} \label{case2}
V_\mathrm{int}(\Phi,\Psi) = - \frac{1}{2} \lambda_2 m^2_\phi \psi^2 \phi^2
\end{equation}

\subsection{Indirect couplings}

These couplings come from $\Phi$ and $\Psi$'s couplings to other fields at the point of enhanced symmetry $\Phi = \Psi = 0$, and are terms of the general form of those in Eq.~(\ref{genrenorm}).
In particular, $\Psi$ and $\Phi$ may couple indirectly through a third field, leading to a $\psi$-dependent renormalization of $\Phi$'s mass
\begin{equation} \label{case3}
V_\mathrm{int}(\Phi, \Psi) = - \frac{1}{2} \alpha_\psi m^2_\phi \ln \left( \frac{\psi}{\psi_*} \right) \phi^2
\end{equation}
$\alpha_\psi$ is expected to be small but not very small since it is a standard renormalisation coefficient derived from $\Phi$'s gauge and Yukawa couplings.
$\psi_*$ is the value of $\psi$ at some convenient time around horizon crossing of cosmologically relevant scales.
Note that in order to make the full potential, Eq.~(\ref{Vfull}), independent of $\psi_*$, we require
\begin{equation}
\phi_* \propto \psi_*^{\alpha_\psi/\alpha_\phi}
\end{equation}
and $\phi_*$ is interpreted as the minimum of $\phi$'s potential at the time when $\psi = \psi_*$.

\section{Background evolution}
\label{background}

Initially, the potential energy $V_0$ drives inflation, $\Phi$ is held at $\phi = \phi_*$ and $\Psi$ starts rolling away from $\Psi=0$.

\subsection{$\Psi$}

Solving
\begin{equation}
\ddot\psi + 3 H_0 \dot\psi - m^2_\psi \psi = 0
\end{equation}
where $H_0 = \sqrt{V_0/3}$ gives
\begin{equation} \label{psi}
\psi \propto a^p
\end{equation}
where $a$ is the scale factor and
\begin{equation} \label{p}
p = \frac{3}{2} \left( \sqrt{1+ \frac{4 m^2_\psi}{3 V_0}} - 1 \right)
\end{equation}
Since $m^2_\psi \sim V_0$, we expect $p$ to be of order one.
Using this, we can more accurately determine the Hubble parameter
\begin{equation} \label{H}
3 H^2 = V_0 - \frac{1}{2} m^2_\psi \psi^2 + \frac{1}{2} \dot{\psi}^2 = V_0 \left( 1 - \frac{1}{2} p \psi^2 \right)
\end{equation}
The contributions of $\Phi$ to the Hubble parameter will not be important.

\subsection{$\Phi$}

\subsubsection{Direct couplings}

Since $\psi \ll 1$, the couplings are a small perturbation to $\Phi$'s potential.
Therefore, to leading order in $\psi$
\begin{equation} \label{phi2eom}
\ddot\phi + 3 H_0 \dot\phi + \alpha_\phi m^2_\phi \left( \phi - \phi_* \right) + \left. \frac{\partial V_\mathrm{int}}{\partial \phi} \right|_{\phi_*} = 0
\end{equation}

\paragraph{Linear coupling}

Solving Eq.~(\ref{phi2eom}) with $V_\mathrm{int}$ given by Eq.~(\ref{case1}), we get
\begin{equation} \label{phicase1}
\phi = \left(1 + \lambda_1 \mu^2_1 \psi \right) \phi_*
\end{equation}
to leading order in $\psi$, where
\begin{equation}
\mu^2_1 = \frac{m^2_\phi}{\left( p^2 + 3 p \right) H_0^2 + \alpha_\phi m^2_\phi}
\end{equation}

\paragraph{Quadratic coupling}

Solving Eq.~(\ref{phi2eom}) with $V_\mathrm{int}$ given by Eq.~(\ref{case2}), we get
\begin{equation} \label{phicase2}
\phi = \left( 1 + \lambda_2 \mu^2_2 \psi^2 \right) \phi_*
\end{equation}
to leading order in $\psi^2$, where
\begin{equation}
\mu^2_2 = \frac{m^2_\phi}{\left( 4 p^2 + 6 p \right) H_0^2 + \alpha_\phi m^2_\phi}
\end{equation}

\subsubsection{Indirect coupling}

In this case, the equation of motion for $\phi$ is
\begin{equation} \label{phi3eom}
\ddot\phi + 3 H_0 \dot\phi + \alpha_\phi m^2_\phi \ln\left( \frac{\phi}{\phi_*} \right) \phi - \alpha_\psi m^2_\phi \ln\left( \frac{\psi}{\psi_*} \right) \phi = 0
\end{equation}
which has solution
\begin{equation} \label{1stordersol}
\phi = \kappa \phi_* \left( \frac{\psi}{\psi_*} \right)^\frac{\alpha_\psi}{\alpha_\phi}
\end{equation}
where
\begin{equation}
\kappa = \exp{\left[ - \left( 3 p + p^2 \frac{\alpha_\psi}{\alpha_\phi} \right) \frac{\alpha_\psi}{\alpha_\phi^2} \frac{H_0^2}{m^2_\phi} \right]}
\end{equation}

\section{Perturbations}
\label{perturbation}

\subsection{General formalism}
\label{perteqns}

\subsubsection{$\delta N$ formalism}
\label{deltaNformalism}

The $\delta N$ formalism \cite{MSEDS,AAS} is a very general, precise and intuitive way of calculating the final curvature perturbation, taking into account the contributions from all the light scalar fields during inflation.
The final curvature perturbation on comoving hypersurfaces is given by the perturbation in $e$-folding number from an initial flat hypersurface to the final comoving one
\begin{equation} \label{deltaNfull}
\mathcal{R}_\mathrm{c}(t_\mathrm{f}) = \delta \mathcal{N}(t_\mathrm{i}, t_\mathrm{f})
\end{equation}
where $t_\mathrm{i}$ is a time soon after cosmological scales leave the horizon during inflation, $t_\mathrm{f}$ is a time after all trajectories have coalesced so that $\mathcal{R}_\mathrm{c}$ is constant on superhorizon scales, and
$\delta\mathcal{N}$ is the perturbed $e$-folding number.

The background $e$-folding number $N = N(\phi,\dot\phi^\mathbf{a},t_\mathrm{f})$, where $\phi$ is a point in the field space of the relevant light scalar fields.
Taking the growing mode of the background dynamics gives $\dot{\phi}^\mathbf{a} = \dot\phi^\mathbf{a}(\phi)$, so we can take $N = N(\phi,t_\mathrm{f})$.
Therefore Eq.~(\ref{deltaNfull}) can be rewritten as
\begin{equation} \label{deltaNreduced}
\mathcal{R}_\mathrm{c}(t_\mathrm{f}) = \mathinner{\frac{\partial N}{\partial \phi^{\mathbf{a}}}(\phi(t_\mathrm{i}),t_\mathrm{f})} \mathinner{\delta\phi^\mathbf{a}(t_\mathrm{i})}
\end{equation}
where $\delta \phi^\mathbf{a}(t_\mathrm{i})$ is the scalar field perturbation on the initial flat hypersurface.
Note that $N(\phi,t_\mathrm{f})$ is a non-local quantity depending on the whole trajectory of evolution.

In our model, Eq.~(\ref{deltaNreduced}) becomes
\begin{equation}
\mathcal{R}_\mathrm{c} = \frac{\partial N}{\partial \psi} \mathinner{\delta\psi} + \frac{\partial N}{\partial \phi} \mathinner{\delta\phi} + \frac{\partial N}{\partial \theta} \mathinner{\delta\theta}
\end{equation}
Soon after cosmological scales leave the horizon
\begin{equation}
\frac{\partial N}{\partial \psi} \delta \psi \sim \left( \frac{H}{\dot{\psi}} \right) H \sim \frac{H}{\psi}
\end{equation}
The positive mass squared in the $\phi$ direction will tend to suppress $\left( \partial N / \partial \phi \right) \delta\phi$. It could still be significant, but for simplicity we do not consider it here.
For small $\phi$, the angular direction $\theta$ is flat so its fluctuations are not suppressed and will be magnified as $\phi$ increases \cite{KKEDS,KKEDS2}. Since
\begin{equation}
\frac{\partial N}{\partial \theta} \delta \theta \sim \frac{\partial N}{\partial \theta} \frac{H}{\phi}
\end{equation}
and $\partial N / \partial \theta \sim \mathcal{O}(1)$ or larger \cite{KKEDS,KKEDS2},
the final curvature perturbation is dominated by the contribution from the angular perturbation as long as $\phi \ll \psi$ when cosmological scales leave the horizon.
Assuming this, the amplitude of the spectrum is given by
\begin{equation}
\mathcal{R}_\mathrm{c} = \frac{\partial N}{\partial \theta} \mathinner{\delta\theta}
\sim \frac{\partial N}{\partial \theta} \frac{H}{\phi}
\end{equation}
which has considerable freedom.
However, $\partial N / \partial \theta$ is independent of time for small $\phi$, and so we just need $\delta\theta$ to determine the scale dependence of the spectrum.

\subsubsection{Equation of motion of $\delta \theta$}
\label{deltaNformalism}

The evolution of $\delta \theta$ is given by \cite{KKEDS}
\begin{equation} \label{eom}
\frac{d}{d t} \left( a^3 \phi^2 \frac{d \, \delta\theta}{d t} \right) + a k^2 \phi^2 \delta \theta = 0
\end{equation}
Defining $\varphi \equiv a \phi \, \delta\theta$ and $x \equiv -k\eta$ where $\eta$ is conformal time, we get
\begin{equation} \label{keyequation}
\frac{d^2 \varphi}{d x^2} + \left( 1 - \frac{2}{x^2} \right) \varphi = \frac{g}{x^2} \varphi
\end{equation}
where
\begin{eqnarray} \label{g}
g & = & \frac{1}{x a \phi} \left[ \frac{d^2 \left(x a \phi \right)}{\left( d \ln{x} \right)^2} - 3 \frac{d \left( x a \phi \right) }{d \ln{x}} \right]
\nonumber \\
& = & \left( \frac{x a H}{k} \right)^2 \left[ \frac{\ddot{\phi}}{H^2 \phi} + 3 \frac{\dot{\phi}}{H \phi} + \frac{\dot{H}}{H^2} + 2 -2 \left( \frac{k}{x a H} \right)^2 \right]
\end{eqnarray}
$\phi = \phi(\psi)$ and from Eq.~(\ref{psi})
\begin{equation}\label{psix}
\psi = \psi_* \left( \frac{x}{x_*} \right)^{-p}
\end{equation}
We can apply the known results of Refs. \cite{EDS,EDSDHL} to calculate $\delta \theta$, depending on the behavior of $g$.
If $d \ln g / d \ln a \ll 1$, the standard slow-roll approximation is valid.
Otherwise the general slow-roll approximation will be applied.

\subsection{Power spectra}
\label{secpowerspectrum}

\subsubsection{Direct couplings}

\paragraph{Linear coupling}

Here we take $H = H_0$.
Then
\begin{equation}
x = \frac{k}{aH_0}
\end{equation}
and
\begin{equation}
g = \left( p^2 + 3p \right) \lambda_1 \mu^2_1 \psi
\end{equation}
Therefore, using the results of Ref.~\cite{EDS}, the power spectrum is
\begin{equation}
P_{\mathcal{R}_\mathrm{c}} = \left( \frac{H_0}{2\pi \phi_*} \frac{\partial N}{\partial \theta} \right)^2
\left\{ 1 - 2 \lambda_1 \mu^2_1 \psi_* \left[ 2^p \cos{\left( \frac{\pi p}{2} \right)} \frac{\Gamma(2-p)}{1+p} \left(\frac{k}{a_*H_0}\right)^p - 1 \right] \right\}
\end{equation}
and the spectral index is
\begin{equation}
n_{\mathcal{R}_\mathrm{c}} - 1 = - 2 p \lambda_1 \mu^2_1 \psi_* \left[ 2^p \cos{\left( \frac{\pi p}{2} \right)} \frac{\Gamma(2-p)}{1+p} \right] \left(\frac{k}{a_*H_0}\right)^p
\end{equation}
for $p < 2$.
These results are basically the same as those of Ref.~\cite{KKEDS}, though the origin is different.

\paragraph{Quadratic coupling}

Here we have terms of order $\psi^2$, and so must use Eq.~(\ref{H}).
Then
\begin{equation} \label{x}
x = \frac{k}{aH} \left[ 1 + \frac{p^2 \psi^2}{2 \left( 1 - 2p \right)} \right]
\end{equation}
and
\begin{eqnarray}
g = \left( 4p^2 + 6p \right) \sigma \psi^2
\end{eqnarray}
where
\begin{equation}
\sigma = \left[ \lambda_2 \mu^2_2 + \frac{p}{4 \left( 1-2p \right)} \right]
\end{equation}
This gives a similar power spectrum
\begin{equation}
P_{\mathcal{R}_\mathrm{c}} = \left( \frac{H_*}{2\pi \phi_*} \frac{\partial N}{\partial \theta} \right)^2
\left\{ 1 - 2 \sigma \psi_*^2 \left[ 2^{2p} \cos(\pi p) \frac{\Gamma(2-2p)}{1+2p} \left(\frac{k}{a_*H_*}\right)^{2p} - 1 \right] - \frac{p^2 \psi_*^2}{1-2p} \right\}
\end{equation}
and spectral index
\begin{equation}
n_{\mathcal{R}_\mathrm{c}} - 1 = - 4 p \sigma \psi_*^2 \left[ 2^{2p} \cos(\pi p) \frac{\Gamma(2-2p)}{1+2p} \right] \left(\frac{k}{a_*H_*}\right)^{2p}
\end{equation}
for $p < 1$.

\subsubsection{Indirect coupling}

Here we take $H = H_0$.
Then
\begin{equation}
x = \frac{k}{aH_0}
\end{equation}
and
\begin{equation} \label{gapprox}
g = p^2 \left( \frac{\alpha_\psi}{\alpha_\phi} \right)^2 + 3p \left( \frac{\alpha_\psi}{\alpha_\phi} \right)
\end{equation}
Therefore, using the results of Ref.~\cite{EDSDHL}, the power spectrum is
\begin{equation}
P_{\mathcal{R}_\mathrm{c}} = \left( \frac{H_0}{2\pi \kappa\phi_*} \frac{\partial N}{\partial \theta} \right)^2
\left[ 2^{p \frac{\alpha_\psi}{\alpha_\phi}} \frac{\Gamma\left(\frac{3}{2}+p\frac{\alpha_\psi}{\alpha_\phi}\right)}{\Gamma\left(\frac{3}{2}\right)} \right]^2 \left( \frac{k}{a_* H_0} \right)^{- 2 \left( p \frac{\alpha_\psi}{\alpha_\phi} \right)}
\end{equation}
and the spectral index is
\begin{equation} \label{index1}
n_{\mathcal{R}_\mathrm{c}} - 1 = - 2 \left( p \frac{\alpha_\psi}{\alpha_\phi} \right)
\end{equation}
Note that the form of the spectrum is completely different from the previous two cases and the results of Ref.~\cite{KKEDS}.

The renormalization coefficients $\alpha_\phi$ and $\alpha_\psi$ depend on $\Phi$ and $\Psi$'s gauge and Yukawa couplings near the point of enhanced symmetry, as well as the spectrum of soft supersymmetry breaking masses, and so have a wide range of possible values.
An order of magnitude or two between their values would be quite natural.
Thus, the observational constraint of an approximately scale invariant spectrum
\begin{equation} \label{basiccondition}
p \alpha_\psi \ll \alpha_\phi
\end{equation}
can be obtained naturally.

\section{Conclusion}
\label{conclusion}

Moduli provide a very simple and natural realization of inflation, making modular inflation very attractive.
However, the usual treatment of moduli for inflation is over simplified and does not produce the observed scale invariant spectrum without fine tuning.

In Ref.~\cite{KKEDS}, Kadota and Stewart considered the nature of moduli more carefully and developed a successful scenario of modular inflation in which a nearly scale invariant spectrum can be obtained naturally.
However, their work was still over simplified, since they did not consider the likely couplings between the moduli.

In this paper, we analyzed the effect of couplings between the moduli on the spectrum of perturbations.
We considered the three simplest generic couplings that might be expected to occur between the moduli: a direct coupling linear in $\Psi$, a direct coupling quadratic in $\Psi$, and an indirect coupling.
The direct couplings both gave
\begin{equation}
\frac{d n}{d \ln k} \propto n - 1
\end{equation}
essentially the same result as Ref.~\cite{KKEDS}.
However, indirect couplings gave a power law spectrum
\begin{equation}
\frac{d n}{d \ln k} \simeq 0
\end{equation}

Thus, one should carefully take into account all couplings expected in realistic models to make robust predictions for the spectrum of perturbations.

\subsection*{Acknowledgements}

EDS thanks Takahiro Tanaka for helpful discussions.
This work was supported in part by
ARCSEC funded by the Korea Science and Engineering Foundation and the Korean Ministry of Science,
the KOSEF grant R01-2005-000-10404-0,
the Korea Research Foundation grant KRF-2005-201-C00006 funded by the Korean Government (MOEHRD),
and Brain Korea 21.

\end{document}